# Perfectly Aligned Shallow Ensemble Nitrogen-Vacancy Centers in (111) Diamond

Hitoshi Ishiwata[1], Makoto Nakajima[2], Kosuke Tahara[2], Hayato Ozawa[2], Takayuki Iwasaki[2] and Mutsuko Hatano[2*]

[1]*Academy of Co-creative Environment, Energy, Education and Science, Tokyo Institute of Technology, Meguro, Tokyo 152-8552, Japan.*

[2]*Department of Electrical and Electronic Engineering, Tokyo Institute of Technology, Meguro, Tokyo 152-8552, Japan.*

We report the formation of perfectly aligned, high-density, shallow nitrogen vacancy (NV) centers on the (111) surface of a diamond. The study involved step-flow growth with a high flux of nitrogen during chemical vapor deposition (CVD) growth, which resulted in the formation of a highly concentrated ($>10^{19}$ $cm^{-3}$) nitrogen layer approximately 10 nm away from the substrate surface. Photon counts obtained from the NV centers indicated the presence of $6.1 \times 10^{15}$-$3.1 \times 10^{16}$ $cm^{-3}$ NV centers, which suggested the formation of an ensemble of NV centers. The optically detected magnetic resonance (ODMR) spectrum confirmed perfect alignment (more than 99 %) for all the samples fabricated by step-flow growth via CVD. Perfectly aligned shallow ensemble NV centers indicated a high Rabi contrast of approximately 30 % which is comparable to the values reported for a single NV center. Nanoscale NMR demonstrated surface-sensitive nuclear spin detection and provided a confirmation of the NV centers depth. Single NV center approximation indicated that the depth of the NV centers was approximately 9-10.7 nm from the surface with error of less than ±0.8 nm. Thus, a route for material control of shallow NV centers has been developed by step-flow growth using a CVD system. Our finding pioneers on the atomic level control of NV center alignment for large area quantum magnetometry.

* Author to whom correspondence should be addressed. Electronic mail:  hatano.m.ab@m.titech.ac.jp

Nitrogen-vacancy (NV) color centers in diamond emerged as a breakthrough material to realize quantum sensing and quantum information processing. These centers possess unique spin-dependent fluorescence combined with microwave coherent manipulation and constitute a material platform for a quantum magnetometer[1,2]. Furthermore, a NV center located in close proximity to an external spin of interest allows for statistical nuclear polarization detection at the nanoscale for magnetic sensing applications[3,4,5]. Nanoscale nuclear magnetic resonance (NMR) allows for small detection volume in the order of approximately (5 nm)[3], and can be utilized for the determination of single protein structure, in contrast to the standard magnetic detection techniques such as NMR and magnetic force microscopy (MFM)[6,7]. A fundamental limitation of an NV center based magnetometer is the material control required to confine the NV center in the vicinity (<10 nm) of the substrate surface with a high magnetic sensitivity. Previous studies that examined shallow NV centers focused on either a high-density ensemble for two-dimensional large area imaging or a single NV center for high contrast and high coherent time to obtain a minimal detection volume using nanoscale NMR. However, it was found necessary to combine spatial localization of a NV centers with alignment, high density, and a long spin coherence time ($T_2$) to obtain high magnetic sensitivity. The alignment of NV centers in an ensemble is the key to accomplish high contrast while maintaining high signal to noise ratio for high magnetic sensitivity with low accumulation time. In this regard, low energy ion implantation is the most common technique utilized for the production of NV centers in the vicinity of a surface[8]. However, this methodology suffers from large depth dispersion (>10 nm) of the NV centers due to ion straggling and channeling effects[9,10]. Additionally, high-density surface defects formed during implantation affect the spin coherence time and the ensembles show a random orientation with this technique[12,13,14]. Existing studies include reports of CVD growth that demonstrated a narrow distribution in the confinement of NV centers in the vicinity of a surface[11] and their atomic alignment on (100), (110), (113), and (111) substrates for the formation of thick diamond films[14,15,16,17,18,19,20]. Nearly all previous studies have focused on either low density NV centers (<$10^{13}$ cm$^-$

$^3$) in the vicinity of a surface with no alignment[21,22,23] or the formation of NV ensembles with alignment in bulk.

In this paper, the formation of a perfectly aligned high-density shallow NV center film for surface-sensitive detection of nuclear spin has been demonstrated. Results obtained from SIMS measurement combined with an effective depth obtained from nanoscale NMR measurement confirms presence of shallow NV center approximately 9-10.7 nm from surface with error of less than ±0.8 nm. The results of this study offer a path toward controlling the alignment of shallow NV center ensembles.

In this study, NV-containing diamond films were grown on diamond IIa (111) substrates by using a microwave plasma chemical vapor deposition (MPCVD) system using $CH_4$ and $H_2$ as source gases. During the growth, $N_2$ gas was introduced as a nitrogen source to form NV centers in the diamond films. A shallow NV center was formed by increasing the time of growth from 30 to 120 s at growth rate of 0.3 nm s$^{-1}$ as confirmed by secondary ion mass spectrometry (SIMS). The growth condition included 75 Torr pressure, 620 W power, 900 °C temperature, and a total gas flow of 1000 sccm with $CH_4$: 0.5 sccm, $N_2$: 3.2–4 sccm and $H_2$ as a carrier gas. An intrinsic diamond was grown for 7 h prior to the formation of the NV centers. NV centers were formed on the surface without intrinsic diamond cap layer. The off-angle of the substrates corresponded to 2-3 ° along the $<\bar{1}\bar{1}2>$ direction and off-direction less than 5 ° from the $<\bar{1}\bar{1}2>$ direction. The morphology of the samples was investigated by atomic force microscopy (AFM). The fluorescence intensity of the NV centers was measured by using a home-built confocal microscope system that was equipped with a 532 nm laser, avalanche photo diode detectors, and a spectrometer. Optically detected magnetic resonance (ODMR) was performed to analyze the alignment ratio of the NV axis. Surface magnetic sensitivity was determined by performing nanoscale NMR using a XY8 pulse sequence.

A typical AFM image of the sample is shown in Figure 1(a). Step-flow is observed towards the $<\bar{1}\bar{1}2>$ direction as indicated by a black arrow in Figure 1(a) and the cross sectional scan of the AFM image in this location is shown in Figure 1(b). The average step height lies in the range of 1–4 nm. The step heights of 1 nm and 4 nm are indicated by arrows in Figure 1(b). Small steps of 1 nm and a considerably smaller terrace size are packed tightly between the larger step height terrace. The step height distribution corresponds to a distribution in nanoscale bunching that occurs during intrinsic layer deposition. Figure 1 (c) shows a confocal XY scan of the sample that illustrates the high emission counts observed throughout the sample. Confocal spot size estimated by 300 nm diameter and thickness obtained from SIMS measurements were used for calculation of the confocal spot volume. Emission counts were compared with photon counts from single NV center for calculation of an ensemble NV density. Photon counts observed in the sample corresponds to the NV centers density of $6.1\times10^{15}$ cm$^{-3}$–$3.1\times10^{16}$ cm$^{-3}$. The observed NV center distribution can be correlated to the distribution of step height in the AFM image, which is caused by difference in the speed of propagation of steps. Higher steps are caused by bunching of step-flow wherein higher steps propagate at slower speed in the lateral direction and in which the NV centers are more localized when compared to a smaller step size with higher propagation speed. Ensembles of NV centers are observed throughout the sample and have a density that exceeds $6.1\times10^{15}$ cm$^{-3}$.

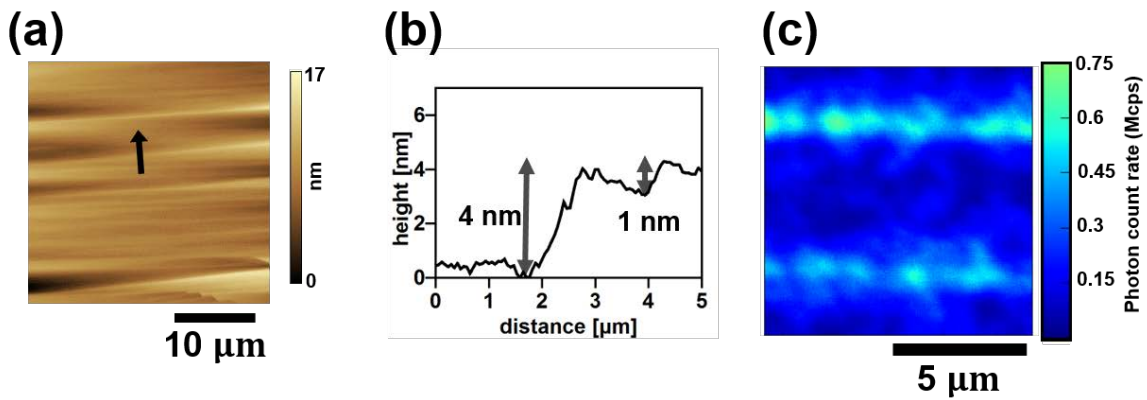

**Fig. 1. Surface morphology and fluorescence characteristics of shallow ensemble NV centers. (a) AFM image (field view of 30 μm × 30 μm) black arrow indicates the $<\bar{1}\bar{1}2>$ direction and the**

**location of the cross sectional scan shown in (b). (b) AFM cross sectional profile. (c) XY confocal scan image (field of view of 10 μm x 10 μm).**

The SIMS measurement confirms the presence of a nitrogen layer ($>10^{19}$ cm$^{-3}$) having variable thickness depending on the different growth times. Figure 2 shows the SIMS measurement performed on a delta doped layer that was obtained after 30, 60, and 120 s of growth time. All samples used for SIMS measurements show full coverage of the substrate with NV centers. These measurements also confirm the background level of nitrogen ($< 5 \times 10^{16}$ cm$^{-3}$) in the intrinsic layer. A variation in the growth time of 30, 60, and 120 s indicates film thicknesses of 12, 18, and 36 nm, respectively, corresponding to a growth rate of 0.29 nm s$^{-1}$. The nitrogen concentration shows an exponential increase and saturation at $3 \times 10^{19}$ cm$^{-3}$ which suggests a difference in the thickness of delta doped layer with a comparable concentration of nitrogen. The presence of a high nitrogen concentration ($>10^{19}$ cm$^{-3}$) in the vicinity of the surface indicates the confinement of delta doped NV centers with an approximate width of 10 nm corresponding to the shortest growth time of 30 s. High concentration of nitrogen ($>10^{19}$ cm$^{-3}$) was induced by combination of high impurity incorporation efficiency on (111) diamond substrate[24] and use of high nitrogen to carbon ratio ranging from 8-6.4.

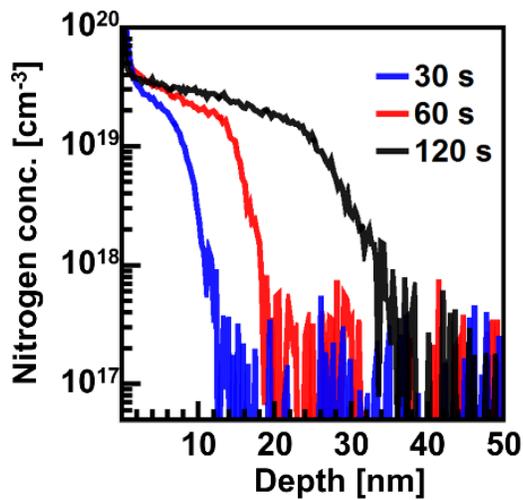

**Fig. 2. SIMS measurement with respect to the nitrogen concentration of shallow ensemble NV centers with growth times of 30 s, 60 s and 120 s. High nitrogen density observed down to depth of 1-2 nm is caused by surface adsorbents and it does not account for actual concentration in film.**

Figure 3 (a) shows the photoluminescence (PL) spectrum of the shallow NV ensembles. A zero phonon line (ZPL) can be clearly observed at 637 nm with the appearance of a phonon side band at a higher wavelength that confirms the presence of a $NV^-$ centers in the films. The $NV^0$ emission that is typically observed at 575 nm is notably absent in the PL spectrum. The diamond Raman peak is observed at 572 nm and a second order Raman peak of a diamond is observed at 612 nm. The PL spectrum also confirms the formation of a stable negatively-charged state within 10 nm from the hydrogen terminated sample surface. This effect is attributed to the high concentration of nitrogen ($> 10^{19}$ cm$^{-3}$) introduced into the film to provide n-type conductivity by stabilizing a Fermi level at an energy exceeding the $NV^-$ ground state energy level[8]. All samples described here were hydrogen terminated during CVD growth. Continuous wave (CW)-ODMR spectrum measured under a static magnetic field along the [111] direction was used to investigate the crystal orientation of the NV centers. This configuration shows the resonance lines from the NV centers along the [111] direction (NV[111]) at $2.87 \pm (\gamma/2\pi)B$ GHz, and along the other three directions at $2.87 \pm(\gamma/2\pi)B \cos(109°)$ GHz in the ODMR spectrum[25]. Magnetic field was first calibrated by using randomly aligned NV centers to confirm peaks for [111] direction and other directions. Figure 3 (b) shows a typical ODMR spectrum of shallow NV ensembles with 25 mT of magnetic field. Two resonance lines are observed for the sample, with no additional two peaks corresponding to a direction separate from the [111] direction. Comparison of root mean square (RMS) value on background noise with the signal obtained from ODMR spectrum shows perfect alignment ratio of more than 99 %. Previous reports[21,22,23] on the fabrication of a shallow NV centers by CVD technique indicated a NV concentration in the approximate range of $10^{12}$~$10^{13}$ cm$^{-3}$. The sample in this study demonstrates more than hundred times increase in the density of the NV centers created in the vicinity of the surface with perfect alignment. The spin coherent property of shallow NV ensembles is evaluated by performing the Rabi oscillation and Hahn echo measurements. Figure 3(c) shows the Rabi oscillation observed on perfectly aligned shallow NV ensembles. These ensembles indicate a high contrast of approximately 30 %, which is comparable to that reported for single NV centers experiments[20]. The high contrast demonstrates the alignment of ensemble NV centers in the vicinity of the surface on the (111) substrate that combines the benefit of a single NV center for high contrast with a high number of NV centers. Spin echo measurements were also performed to determine the spin relaxation time ($T_2$) in the shallow NV samples. Figure 3 (d) is fitted to the exponential decay function in order to evaluate $T_2$, which was found to be 6 µs for the sample that was grown for 30 s under a $N_2$ flow rate of 3.2 sccm. These results indicate a slight increase in the value of $T_2$; with variations in the nitrogen flow rate during CVD growth from 4 sccm to 3.2 sccm, the $T_2$ value changed from approximately 4 µs to 6 µs. The coherent time of 6 µs is attributed to the high-density P1 center which could also be confirmed from the SIMS measurements that showed a nitrogen concentration exceeding $10^{19}$ cm$^{-3}$ [26,27]. No distribution in $T_2$ values were observed between NV centers created on the step edge or step terrace.

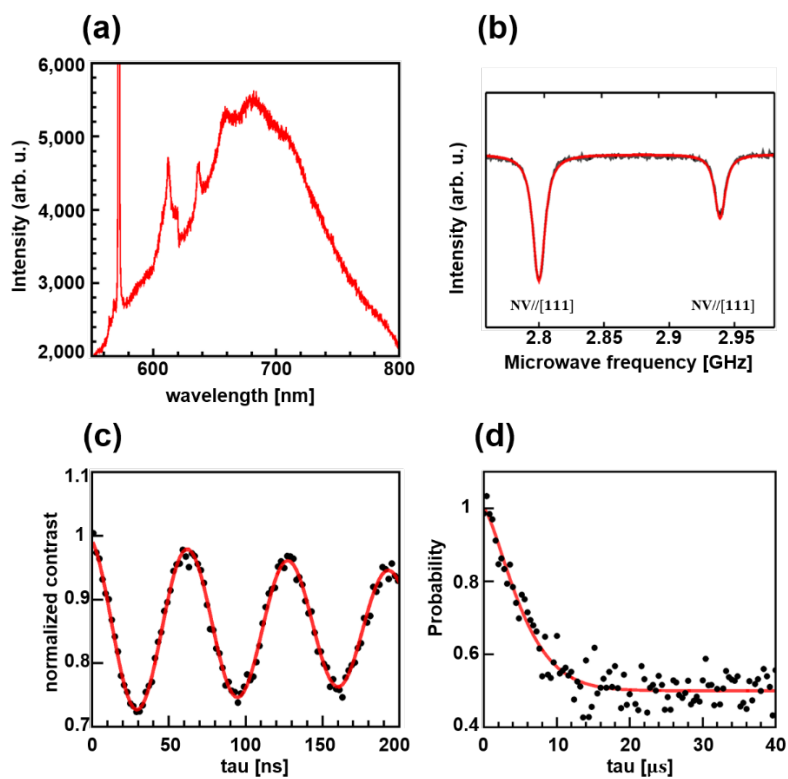

**Fig. 3.** (a) PL spectrum of shallow ensemble NV centers (b) ODMR spectrum obtained from shallow ensemble NV centers illustrating their perfect alignment. Magnetic field of 25 mT was applied for measurement. While the black line shows the actual results of the ODMR spectrum, the red curve is a Voigt function fitted result. Peaks are indicated with a corresponding direction ([111]) of the NV centers. Spin coherence measured by (c) Rabi oscillation measurement and (d) spin echo measurement. Black circles indicate the experimental results and the red curves indicate best fit.

The surface magnetic sensitivity and depth of the NV centers are demonstrated by performing nanoscale NMR using a XY8 sequence. With respect to the nanoscale NMR, statistical polarization of $^1$H and $^{19}$F spin is detected at Larmor frequency, which is determined by an applied DC magnetic field. A periodic sequence of microwave $\pi$ pulses is applied during coherent evolution in order to dynamically decouple the environmental magnetic noise from the NV centers. The local magnetic field is measured by the ODMR spectrum obtained from the shallow NV centers ensemble. The detection of the nuclear spin is confirmed by a sweeping pulse spacing $\tau$, and Larmor precession is detected when $\tau$ matches half the periodicity of the precession. The XY8-80 (total of 80 pulse) sequence was performed at three different values of the magnetic field in order to obtain a change in the contrast at a frequency that corresponded to the Larmor frequency of fluorines in Fomblin Y HVAC 140/13 oil and thin layer of proton reported previously[4]. A change in contrast was observed with respect to magnetic fields of 24.1, 25.9 and 27.5 mT, and this caused phase accumulation in the NV centers at frequencies corresponding to 1.03, 1.1 and 1.17 MHz for proton spins and 0.97, 1.04 and 1.1 MHz for nuclear spin of fluorines. Figure 4 (b) shows the experimentally observed frequency with respect to changes in the magnetic field. Frequency of protons and fluorines are measured with less than 0.8 % difference from theoretical value. The gyromagnetic ratio of the observed nuclear spin can be obtained from the straight line fit since it shows the dependence of frequency on the applied magnetic field. The slope of straight line fit corresponds to 42.4 MHz/T for proton and 39.9 MHz/T for fluorine. Gyromagnetic ratio obtained from these slopes indicate less than 0.4 % difference in the frequency when compared with a gyromagnetic ratio of 42.58 MHz/T of $^1$H proton and 40.03 MHz/T of fluorine $^{19}$F. A small number of pulses that are applied during the measurements limits the full width at half maximum (FWHM) value obtained from the measurement to 25 kHz. The effect of broadening can be suppressed by increasing $T_2$ and applying a larger number of pulses or via correlation spectroscopy to achieve a resolution limited by $T_1$ that is considerably longer than $T_2$[4,28,29].

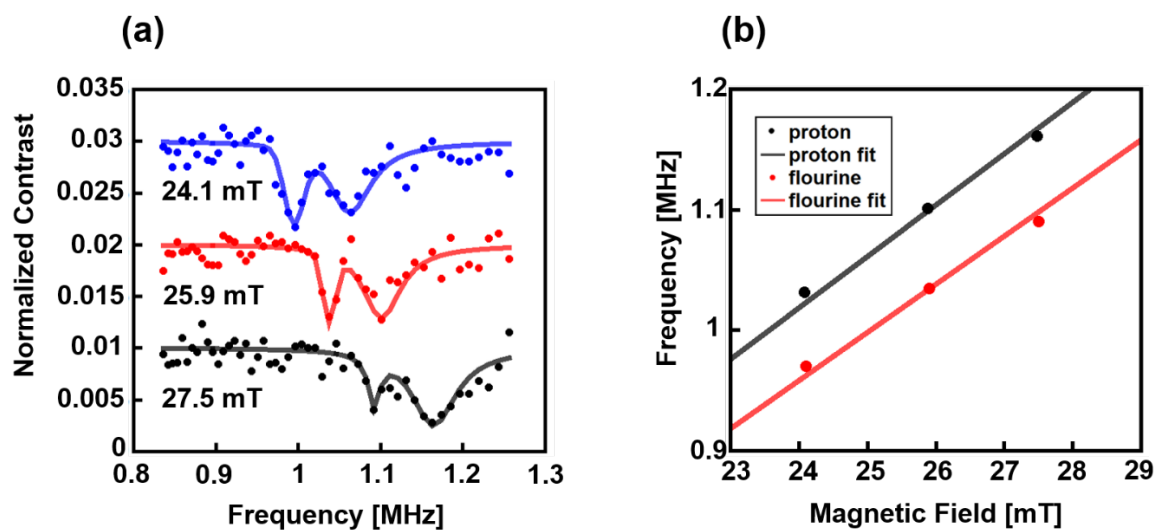

Fig. 4. (a) Normalized contrast obtained from nanoscale NMR from proton and fluorine at different magnetic fields. (b) Frequency of the detected AC magnetic field plotted with respect to the applied DC magnetic field.

The depth of the NV center can be determined by single NV center approximation using methods described previously by DeVience et al[4]. The fitting of the contrast obtained from XY8 sequence calculates the total magnetic field detected from the $^1$H thin layer of proton with a reported hydrogen density of $\rho$= 60 hydrogen atoms per nm$^{-3}$ and the $^{19}$F contained in Fomblin oil with a reported fluorine density of $\rho$= 40 fluorine atoms per nm$^{-3}$. The contrast obtained from the nanoscale NMR measurement is normalized with the decay of the $T_2$ function. After normalization, results obtained from XY8-80 measurements were used for fitting using function described in previous report[4,30] with NV center depth and thickness of proton layer as a fitting parameter. Fitting of our result shows proton layer of approximately 0.7-1.6 nm thickness with error of less than ±0.5 nm and NV centers depth of approximately 9-10.7 nm with error of less than ±0.8 nm. Thickness of proton layer estimated from our study is comparable to the value reported previously[4]. Further study to investigate depth distribution of $T_2$ and depletion of NV centers within 10 nm from surface will be our future work for improving sensitivity of nanoscale NMR.

In conclusion, this study demonstrates that a highly aligned high-density shallow NV centers ensemble is formed by step-flow growth using MPCVD growth on (111) substrates. More than $6.1\times10^{15}$ cm$^{-3}$–$3.1\times10^{16}$ cm$^{-3}$ NV centers are detected from confocal scan. The results demonstrate highest NV density in the vicinity of the surface with perfect alignment of more than 99 %. Surface sensitive magnetic field measurement was performed by observing thin layer of proton and fluorine contained in Fomblin oil by nanoscale NMR using XY8-80 pulse sequence. The single NV center approximation indicates that the depth of the NV centers is approximately 9-10.7 nm from surface with error of less than ±0.8 nm. Our finding offers a route for material engineering for future of quantum magnetometry using NV centers that requires atomic level control of NV centers alignment for precise alignment of magnetic field and surface sensitive magnetic field detection in nanoscale for wide field imaging.

**Acknowledgments**

This work was supported by CREST, JST.

1. F. Jelezko, T. Gaebel, I. Popa, A. Gruber, and J. Wrachtrup, Phys. Rev. Lett. 92, 076401 (2004).

2. G. Balasubramanian, I. Y. Chan, R. Kolesov, M. Al-Hmoud, J. Tisler, C. Shin, C. Kim, A. Wojcik, P. R. Hemmer, A. Krueger, T. Hanke, A. Leitenstorfer, R. Bratschitsch, F. Jelezko, and J. Wrachtrup, Nature **455**, 648 (2008).

3. J. M. Taylor, P. Cappellaro, L. Childress, L. Jiang, D. Budker, P. R. Hemmer, A. Yacoby, R. Walsworth, and M. D. Lukin, Nat. Phys. **4**, 810 (2008).

4. S. J. DeVience, L. M. Pham, I. Lovchinsky, A. O. Sushkov, N. Bar-Gill, C. Belthangady, F. Casola, M. Corbett, H. Zhang, M. Lukin, H. Park, A. Yacoby, and R. L. Walsworth, Nat. Nano. **10**, 129 (2015)

5. T. Staudacher, F. Shi, S. Pezzagna, J. Meijer, J. Du, C. A. Meriles, F. Reinhard, and J. Wrachtrup, Science **339**, 561 (2013)

6. I. Lovchinsky, A. O. Sushkov, E. Urbach, N. P. de Leon, S. Choi, K. De Greve, R. Evans, R. Gertner, E. Bersin, C. Müller, L. McGuinness, F. Jelezko, R. L. Walsworth, H. Park, and M. D. Lukin, Science **351**, 836-841 (2016)

7. A. Ajoy, U. Bissbort, M. D. Lukin, R. L. Walsworth, and P. Cappellaro, Phys. Rev. X **5**, 011001 (2015)

8. M. V. Hauf, B. Grotz, B. Naydenov, M. Dankerl, S. Pezzagna, J. Meijer, F. Jelezko, J. Wrachtrup, M. Stutzmann, F. Reinhard, and J. A. Garrido, Phys. Rev. B **83**, 81304 (2011).

9. M. S. Grinolds, P. Maletinsky, S. Hong, M. D. Lukin, R. L. Walsworth, and A. Yacoby, Nat. Phys. **7**, 687 (2011).

10. F. Oliveira, S. Momenzadeh, D. Antonov, H. Fedder, A. Denisenko, and J. Wrachtrup, Phys. Status Solidi A **213**, 2044-2050 (2016)

11. Maneesh Chandran, Shaul Michaelson, Cecile Saguy, and Alon Hoffman, Appl. Phys. Lett. **109**, 221602 (2016)

12. Y. Romach, C. Muller, T. Unden, L. J. Rogers, T. Isoda, K. M. Itoh, M. Markham, A. Stacey, J. Meijer, and S. Pezzagna, Phys. Rev. Lett. **114**, 17601 (2015).

13. T. Yamamoto, T. Umeda, K. Watanabe, S. Onoda, M. L. Markham, D. J. Twitchen, B. Naydenov, L. P. McGuinness, T. Teraji, S. Koizumi, F. Dolde, H. Fedder, J. Honert, J. Wrachtrup, T. Ohshima, F. Jelezko, and J. Isoya. Phys. Rev. B **88**, 075206

14. T. Fukui, Y. Doi, T. Miyazaki, Y. Miyamoto, H. Kato, T. Matsumoto, T. Makino, S. Yamasaki, R. Morimoto, N. Tokuda, M. Hatano, Y. Sakagawa, H. Morishita, T. Tashima, S. Miwa, Y. Suzuki, and N. Mizuochi, Appl. Phys. Express **7**, 055201 (2014).

15. V. M. Acosta, E. Bauch, M. P. Ledbetter, C. Santori, K.-M. C. Fu, P. E. Barclay, R. G. Beausoleil, H. Linget, J. F. Roch, F. Treussart, S. Chemerisov, W. Gawlik, and D. Budker, Phys. Rev. B **80**, 115202 (2009).

16. A. M. Edmonds, U. F. S. D'Haenens-Johansson, R. J. Cruddace, M. E. Newton, K.-M. C. Fu, C. Santori, R. G. Beausoleil, D. J. Twitchen, and M. L. Markham, Phys. Rev. B **86**, 035201 (2012).

17. M. Lesik, T. Plays, A. Tallaire, J. Archard, O. Brinza, L. William, M. Chipaux, L. Toraille, T. Debuisschert, A. Gicquel, J. F. Roch, and V. Jacques, Diam. Relat. Mater. **56**, 47 (2015).

18. J. Michl, T. Teraji, S. Zaiser, I. Jakobi, G. Waldherr, F. Dolde, P. Neumann, M. W. Doherty, N. B. Manson, J. Isoya, and J. Wrachtrup, Appl. Phys. Lett. **104**, 102407 (2014).

19. M. Lesik, J.-P. Tetienne, A. Tallaire, J. Achard, V. Mille, A. Gicquel, J.-F. Roch, and V. Jacques, Appl. Phys. Lett. **104**, 113107 (2014).

20. H. Ozawa, K. Tahara, H. Ishiwata, M. Hatano, and T. Iwasaki, APEX **10**, 045501 (2017)

21. K. Ohashi, T. Rosskopf, H. Watanabe, M. Loretz, Y. Tao, R. Hauert, S. Tomizawa, T. Ishikawa, J. Ishi-Hayase, S. Shikata, C. L. Degen, and K. M. Itoh, Nano Lett. **13**, 4733 (2013).

22. K. Ohno, F. Joseph Heremans, L. C. Bassett, B. A. Myers, D. M. Toyli, A. C. Bleszynski Jayich, C. J. Palmstrøm, D. D. Awschalom, Appl. Phys. Lett. **101**, 82413 (2012).

23. C. Osterkamp, J. Lang, J. Scharpf, C. Muller, L. P. McGuinness, T. Diemant, R. J. Behm, B. Naydenov, and F. Jelezko, Appl. Phys. Lett. **106**, 113109 (2015)

24. J Hiromitsu Kato, Toshiharu Makino, Satoshi Yamasaki and Hideyo Okushi. Phys. D: Appl. Phys. 40 (2007) 6189-6200

25. K. Tahara, H. Ozawa, T. Iwasaki, N. Mizuochi, and M. Hatano, Appl. Phys. Lett. **107**, 193110 (2015)

26. Z.-H. Wang and S. Takahashi, Phys. Rev. B **87**, 115122 (2013)

27. L. Rondin, J.-P. Tetienne, T. Hingant, J.-F. Roch, P. Maletinsky, and V. Jacques, Rep. Prog. Phys. **77**, 056503 (2014)

28. A. Laraoui, F. Dolde, C. Burk, F. Reinhard, J. Wrachtrup, C. Meriles, Nat. Commun. **4**, 1651 (2013)

29. P. Kehayiasx, A. Jarmolax, N. Mosavian, I. Fescenko, F. M. Benito, A. Laraoui, J. Smits, L. Bougas, D. Budker, A. Neumann, S. R. J. Brueck, and V. M. Acosta, arXiv:1701.01401

30. L. M. Pham, S. J. DeVience, F. Casola, I. Lovchinsky, A. O. Sushkov, E. Bersin, J. Lee, E. Urbach, P. Cappellaro, H. Park, A. Yacoby, M. Lukin, and R. L. Walsworth, Phys. Rev. B **93**, 045425 (2016)